\documentclass{article}
\pdfoutput=1
\PassOptionsToPackage{numbers, compress}{natbib}

    \usepackage[final]{wdcs2020}
\usepackage{subcaption}

\usepackage[utf8]{inputenc} 
\usepackage[T1]{fontenc}    
\usepackage{hyperref}       
\usepackage{url}            
\usepackage{booktabs}       
\usepackage{amsfonts}       
\usepackage{nicefrac}       
\usepackage{microtype}      
\usepackage{enumerate}
\usepackage{graphicx}
\usepackage{xcolor}
\usepackage[toc,page]{appendix}

\newcommand{\eat}[1]{}
\title{Open4Business (O4B): An Open Access Dataset for Summarizing Business Documents}

\author{%
  Amanpreet Singh\thanks{Work done as an intern for SS\&C Intralinks.} \\
  Department of Computer Science\\
  Stony Brook University\\
  Stony Brook, NY 11790 \\
  \texttt{amanpsingh@cs.stonybrook.edu} \\
   \And
   Niranjan Balasubramanian\\
   Department of Computer Science\\
   Stony Brook University\\
  Stony Brook, NY 11790\\
   \texttt{niranjan@cs.stonybrook.edu} \\
}

\begin{document}

\maketitle

\begin{abstract}
A major challenge in fine-tuning deep learning models for automatic summarization is the need for large domain specific datasets. One of the barriers to curating such data from resources like online publications is navigating the license regulations applicable to their re-use, especially for commercial purposes. As a result, despite the availability of several business journals there are no large scale datasets for summarizing business documents. In this work, we introduce Open4Business (O4B), a dataset of 17,458 open access business articles and their reference summaries. The dataset introduces a new challenge for summarization in the business domain, requiring highly abstractive and more concise summaries as compared to other existing datasets. Additionally, we evaluate existing models on it and consequently show that models trained on O4B and a 7x larger non--open access dataset achieve comparable performance on summarization. We release the dataset, along with the code  \footnote{\url{https://github.com/amanpreet692/Open4Business}} which can be leveraged to similarly gather data for multiple domains.
\end{abstract} 

\section{Introduction}
Generating effective summaries of documents is critical in business and corporate finance. Successful transactions in these domains often require timely processing of hundreds of long, complex and diverse documents. For instance, the due-diligence process in mergers and acquisitions requires the involved parties to go through tons of paperwork \citep{howson2003due}. Concise summaries can save time and effort spent on reading entire documents. 

However, it is challenging to obtain in-domain training data for fine-tuning automatic summarization systems. The documents are often proprietary and confidential or the access to the data itself could be ephemeral. Even with the necessary access, the effort needed for an expert to create reference summaries would be enormous. One way to circumvent these issues is to gather publicly accessible data such as articles from business journals. While most journals are published under licenses that prohibit their commercial use, in recent years there has been a steady increase in open access journals which have more lenient license terms with regards to copyrights and their re-use, even commercially.

This work makes the following contributions to further research and quick prototyping of summarization systems for business documents:
\begin{enumerate}[i)]
    \item O4B - a dataset consisting of 17,458 open access business articles with reference summaries.
    \item Evaluation of current SOTA abstractive models on O4B. We also compare it extensively with other summarization datasets and another huge dataset which is not open access.
    \item An open source implementation which can be used for dataset curation in other domains as well. 
\end{enumerate}

\section{Related work}
Many widely-used summarization datasets mainly cover news articles \citep{see2017get, narayan2018don, grusky2018newsroom}, with the exception of some that cover specific domains such as legislation \citep{eidelman2019billsum} and intellectual property \citep{sharma2019bigpatent}. \citet{cohan2018discourse} previously used ArXiv and PubMed research papers to generate abstracts of long documents. In the financial domain, \citet{el2019multiling} released the Annual Reports of around 4000 companies listed on London Stock Exchange as part of a MultiLing 2019 shared task. Since Annual Reports are generally over 60 pages in length, the task was geared towards generating longer summaries of about 1000 words, which is computationally expensive for the current attention-based models. 

Existing literature that align most closely with O4B, in terms of content and the curation process are \citet{Ammar2018ConstructionOT} and S2ORC \citep{lo-wang-2020-s2orc}. Both consist of a huge volume of article records in multiple domains including business and finance which can be useful for several tasks. 
Even though the former allows commercial use, it doesn't have full text of the articles. Contrastingly, S2ORC has full text but prohibits commercial re-use as it was released under the CC BY-NC license (See Appendix~\ref{sec:background} for more details).
Moreover S2ORC has not been used for benchmarking in summarization as of this writing. We attempt to overcome these shortcomings by providing the article abstract and full text without restrictions on re-use as part of O4B. We also benchmark O4B and compare it to a subset of S2ORC consisting of only business articles, referred to as S2ORCB in the evaluation section.

\section{Open4Business dataset}
We introduce a new summarization dataset, Open4Business (O4B), by parsing open access journal articles from the business domain to obtain their full text and abstract. Only open source tools are used for gathering the data to encourage re-use, and to maintain consistency with the open access theme of this work. \autoref{summ-sample} shows a sample summary from O4B. 

\subsection{Dataset curation}    

O4B is created from a collection of business articles obtained from multiple publishers. We filtered this collection to only those articles which are open access, retrieved and converted their full-text PDF into a structured XML format, and extracted the abstracts as reference summaries. We detail each of these steps below and outline the number of data samples through these stages in Appendix~\ref{sec:output_stages}. 

\paragraph{Selecting open access articles:} To identify open access articles, we used the ISSN of business and financial journals obtained from ISSN-GOLD-OA records \citep{bruns2019issn}, as well as a short list of related keywords (e.g. business, finance, and entrepreneur) and cross-referenced those with Crossref \citep{crossref}, a large scale metadata repository for published articles. Using an existing implementation of Crossref API\footnote{\url{https://github.com/fabiobatalha/crossrefapi}}, we queried the Crossref REST service to retrieve Document Object Identifier (DOI), license details and other article metadata. We process the resulting list to only include articles with open access licenses, specifically those of type 'CC BY (3.0/4.0)'.

\paragraph{Retrieving article PDFs:}
Given the collection of relevant DOIs, we used Unpaywall \citep{piwowar2018state}, another repository that provides the best open access resource locations for a given article. We queried it using Unpywall\footnote{\url{https://github.com/unpywall/unpywall}} and then retrieved the PDF versions of the articles. Out of the 21,397 DOIs, we were able to retrieve PDFs for 18,670. In some cases, there was no DOI record or there were connectivity issues, while for others the DOIs were pointing to non-business journals. 

\paragraph{XML conversion and post-processing:}
We used GROBID \citep{GROBID} to convert PDFs to a structured XML format. 
We have considered only English-language articles that have abstracts. 
For each article we retained only the main text body of the article and removed section headings, tables, figures, citations, and other bibliographic references. 
The resulting collection contains the text of 17,458 articles along with a combination of their titles and abstracts as the gold summaries.

\subsection{Dataset characteristics}
\begin{figure}[!tbp]
  \captionsetup{justification=centering}
  \begin{minipage}[b]{0.4\textwidth}
    \raggedleft
    \includegraphics[scale=0.74]{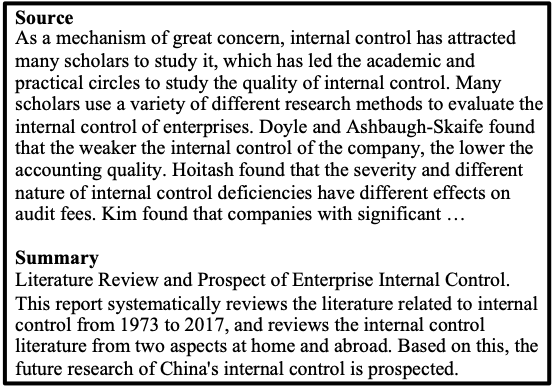}
    \caption{Summary sample from O4B}
    \label{summ-sample}
  \end{minipage}
  \hspace{1.4cm}
  \begin{minipage}[b]{0.4\textwidth}
    \raggedleft
    \includegraphics[scale=0.54]{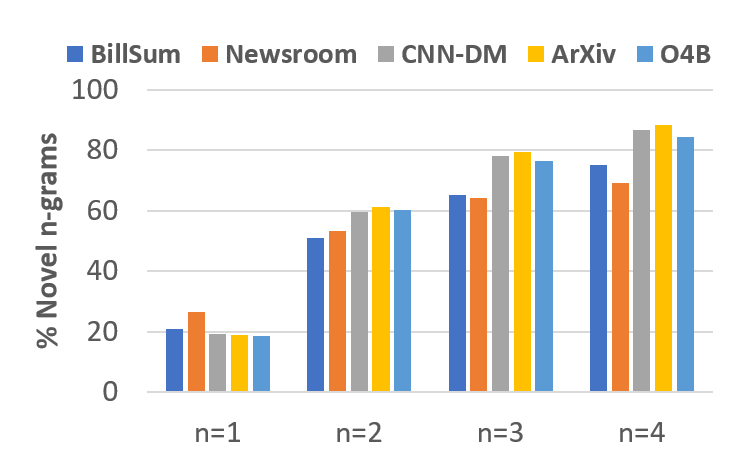}
    \caption{Comparison of unique summary n-grams}
    \label{n-grams}
  \end{minipage}
\end{figure}

To get a measure of the conciseness and abstractiveness of the summaries in O4B, we compare it to a set of benchmark datasets \citep{see2017get, grusky2018newsroom, eidelman2019billsum, cohan2018discourse} from various domains. Following \citep{sharma2019bigpatent}, we compare the reference summaries against their sources in terms of average length in words (to measure conciseness) and unique n-grams (to measure abstractiveness), respectively. 
As shown in \autoref{comp-ratio}, O4B has the highest compression ratio indicating highly compact summaries. The unique summary n-gram distributions shown in \autoref{n-grams} is comparable to ArXiv and CNN-DailyMail(CNN/DM) datasets, which are highly abstractive in nature. Also it has fewer extractive fragments than BillSum and Newsroom. As we show in the next section, our benchmarking experiments indicate that pre-training or fine-tuning on other datasets is not enough for obtaining high scores on O4B, highlighting the need for this dataset.

\begin{table}
	\caption{Comparison of Open4Business (O4B) with other summarization datasets}
	\label{comp-ratio}
	\centering
	\begin{tabular}{lrccc}
	\toprule
Dataset &
  \multicolumn{1}{l}{Total Docs} &
  \multicolumn{1}{c}{\begin{tabular}[c]{@{}c@{}}Mean Words\\ Doc\end{tabular}} &
  \multicolumn{1}{c}{\begin{tabular}[c]{@{}c@{}}Mean Words\\ Summary\end{tabular}} &
  \multicolumn{1}{c}{\begin{tabular}[c]{@{}c@{}}Compression\\ Ratio\end{tabular}} \\
  \midrule
BillSum      & 22,218    & 1,240.21 & 181.98 & 6.81           \\
Newsroom     & 1,145,804 & 684.51   & 29.17  & 23.46          \\
CNN/DM       & 311,971   & 707.65   & 49.66  & 14.25          \\
ArXiv        & 215,500    & 5219.54  & 239.27 & 21.81          \\
\textbf{O4B} & 17,458     & 4764.89  & 173.58 & \textbf{27.45}
\end{tabular}
\end{table}

\section{Evaluation}
Since the focus of this work is solely on abstractive summarization, SOTA encoder-decoder transformer based models \citep{fabbri2020summeval} are used for benchmarking O4B. PyTorch 1.5 implementations of Text-to-Text Transfer Transformer (T5) \citep{raffel2020exploring} and a distilled version of BART \citep{lewis2019bart} (dBART) provided by \citep{Wolf2019HuggingFacesTS} are used for evaluation. Off-the-shelf (base) variants of both the models are fine-tuned on a single 16 GB GPU setup for 2 epochs with a batch size of 2 for consistent comparisons\footnote{Fine-tuning larger models for longer will likely yield better scores but here we only present basic bench-marking results.}. 
For all benchmarking experiments, both the models are fine-tuned on 512 and 1024 tokens each from the source text.

\subsection{ROUGE-N benchmarking}
The O4B benchmark ROUGE-(1/2/L) scores are shown in \autoref{rouge-stats}. 
\begin{table}
            \caption{ROUGE-N Metrics, RG: ROUGE, O4B: Open4Business}
            \label{rouge-stats}
            \centering
            \begin{tabular}{crrrr}
              \toprule
              Architecture     & Tokens     & RG-1 & RG-2 & RG-L \\
             \midrule
                  T5-base              & 512  & 24.30          & 7.21           & 16.33          \\
                  T5-O4B               & 512  & 42.50          & 14.88          & 24.62          \\
                                       & 1024 & 45.34          & 17.39          & 26.31           \\
                  dBART-base           & 512  & 28.46          & 8.10          & 17.89          \\
                  dBART-O4B            & 512  & 43.55          & 15.95          & 25.25          \\
                                       & 1024 & \textbf{45.98} & \textbf{17.90} & \textbf{26.59}\\
              \bottomrule
            \end{tabular}
          \end{table}
Benchmark scores from both the models without any fine-tuning are considered as baselines. During fine-tuning, the models with the best ROUGE-2 score on the dev set are saved as checkpoints. Both the models perform better when they witness a longer portion of the input. But, this comes at cost of longer training times, which increases proportionally with the input length. Overall, the low performance numbers for T5 and dBART - specially for ROUGE-2, indicates that the task of long document summarization is challenging by itself, which is consistent with the findings in \citep{cohan2018discourse}, \citep{zhang2019pegasus} on the ArXiv and PubMed datasets. Sample summaries generated by both the models are presented in Appendix~\ref{sec:gen_result}.
\subsubsection{ Error analysis} 
We observe two main types of errors: word-level repetition, and hallucinating content that is not present in or supported by the input. 
 Consider the sample: \textit{The results show that a mixture of global and local smoothing improves non-parametric estimation, while the local parts allow for the desired flexibility to detect local features.} The source article is about financial data-modeling with sparse mentions of non-parametric estimation but the model combines this information by lifting some text from the source and ignoring other essential parts. Here, the original text is - \textit{Those mixtures are interesting because the global parts can borrow the strength from a larger sample and have a smoothing effect, while the local parts allow for the desired flexibility to detect local features.}

\subsection{Transfer-learning tests with S2ORCB}
To validate the effectiveness of both O4B and open access articles, we conduct transfer-learning experiments. First, we fine-tune T5 and dBART models on O4B and test on the S2ORCB dataset, a large collection of over 119k summary/article pairs. S2ORCB is described in detail in Appendix~\ref{sec:s2orcb}. Here, we consider 1024 source text tokens since that resulted in better benchmark scores. Given the datasets, a sizeable difference is observed in the training times as well. Results in \autoref{train-compare} show that the models fine-tuned on O4B (T5-O4B and dBART-O4B) achieve comparable performance to corresponding models fine-tuned on the S2ORCB dataset. Second, we reverse the setup and fine-tune models on the training portions of S2ORCB and then test on O4B. \autoref{test-compare} indicates a similar gap in performance between the transfer models versus the ones trained on the target dataset (O4B in this case). The gaps hint at the challenges in the summarization task on the O4B dataset that cannot be solved by simply training on a larger dataset. In spite of the huge disparity in dataset sizes, similar model performances indicate that O4B can be effectively utilized for summarization systems. 
       
\begin{table}
 \caption{Transfer learning experiment results, RG: ROUGE}
     \label{tab:transfer-learning}
    \begin{subtable}[h]{0.45\textwidth}
        \centering
        \caption{Performance on S2ORCB test set}
      \label{train-compare}
      \centering
      \begin{tabular}{lrrll}
        \toprule
        Model     & RG-1 & RG-2 & RG-L \\
       \midrule
            T5-O4B  &
            \multicolumn{1}{r}{45.52}          & 17.85          & 26.83          \\
            T5-S2ORCB   & \multicolumn{1}{r}{\textbf{45.59}} & \textbf{18.17} & \textbf{27.32} \\
            dBART-O4B & 46.11                              & 18.26          & 27.04          \\
            dBART-S2ORCB & \textbf{47.36}                     & \textbf{19.47}  & \textbf{28.26}         \\
        \midrule
        Mean Difference  &  0.66    &   0.60    &  0.855    \\
        \bottomrule
      \end{tabular}
    \end{subtable}
    \hfill
    \begin{subtable}[h]{0.45\textwidth}
        \centering
        \caption{Performance on O4B test set}
      \label{test-compare}
      \centering
      \begin{tabular}{lrrll}
        \toprule
        Model     & RG-1 & RG-2 & RG-L \\
       \midrule
            T5-O4B  &   \textbf{45.34}                     & \textbf{17.39} & \textbf{26.31}           \\
            T5-S2ORCB    & 44.04                              & 16.58          & 25.67  \\
            dBART-O4B   & \multicolumn{1}{l}{\textbf{45.98}} & \textbf{17.90} & \textbf{26.59} \\
            dBART-S2ORCB    & \multicolumn{1}{l}{45.72}          & 17.79          & 26.35         \\
        \midrule
        Mean Difference  &  0.78    &   0.46    &  0.44    \\     
        \bottomrule
      \end{tabular}
     \end{subtable}
    
\end{table}

\section{Conclusions} Automatic summarization in business domain is hampered by the lack of access to large scale datasets. Thus, we introduce Open4Business(O4B), a new dataset that consists of open access articles and their abstracts. This presents a new challenge for summarization systems which we hope spurs further research and development of new methods for this domain.
\section*{Acknowledgments}
We would like to thank Prakash Kanchinadam and also the legal team at SS\&C Intralinks for the fruitful discussions and their administrative support in this endeavor.
We thank all anonymous reviewers for their feedback.

\appendix
\section{Appendices}

\subsection{Background}
Traditionally, most of the journals have published under licenses which restrict their re-use in some manner eg. CC BY-NC\footnote{Creative Commons-NonCommercial.}, or prohibit mining them. 
The Open Access movement, which has gained popularity in recent years, allows peer-reviewed articles to be made available immediately for public without any copyright restrictions. Several levels of open access publishing exist with each having its own set of regulations. The readers are encouraged to read about these as it is essential to be aware of any potential restrictions  on re-use of what seems to be public data. Consequently, only the the Gold Open Access level articles licensed as CC BY \footnote{\url{https://creativecommons.org/licenses/by/4.0/}} have been used to create this Open4Business(O4B).
\label{sec:background}

\subsection{O4B through curation stages}
O4B is built from open access business journals published by publishers such as Scientific Research Publishing (SCIRP) \footnote{\url{https://www.scirp.org/}} and Multidisciplinary Digital Publishing Institute (MDPI) \footnote{\url{https://www.mdpi.com/}}. Through every step in the dataset curation process as explained in the main paper, several potential data samples are filtered out due to processing issues as shown in \autoref{dataset-stages}. After final post-processing, remaining articles were split 80/10/10 to have a training, dev and test set of 13,966, 1,746 and 1,746 articles respectively.
\begin{table}[h]
      \caption{Data samples throughout pipeline stages}
      \label{dataset-stages}
      \centering
      \begin{tabular}{lc}
        \toprule
        Step     & Number of samples \\
       \midrule
            License Metadata Filtering (Crossref) & 21,397 \\
            Open Access URL Retrieval (Unpaywall) & 20,432 \\
            PDF Mining                            & 18,670 \\
            XML Conversion (GROBID)               & 18,544 \\
            Final Post-processing                 & \textbf{17,458}\\
        \bottomrule
      \end{tabular}
    \end{table}
\label{sec:output_stages}    

\subsection{Deriving S2ORCB from S2ORC}
S2ORCB is derived from S2ORC by including only those articles that belong to the business domain. S2ORC has the metadata for about 1.8 million business articles but full text is not available for all of them. Thus, the records were then filtered to include only those which had both the article abstracts and full body text. After parsing the text, it is subject to similar post-processing as O4B and any bibliographical references are subsequently removed. Finally, S2ORCB is obtained with article full text as source, and a combination of title and abstract as the reference summary similar to O4B. S2ORCB is split with 95926, 11991 and 11991 articles in the train, dev and test set respectively. \autoref{s2orc-stats} shows the mean word count in source and summary for each of the splits in S2ORCB.
    \begin{table}[h]
          \caption{S2ORC-Business(S2ORCB) Statistics}
          \label{s2orc-stats}
          \centering
          \begin{tabular}{llc}
            \toprule
            Split     & Description     & Mean word count \\
           \midrule
           Train      & Source  & 4611 \\
                  & Summary & 186  \\
           Dev & Source  & 4557 \\
                      & Summary & 187  \\
           Test       & Source  & 4606 \\
                      & Summary & 186 \\
            \bottomrule
          \end{tabular}
        \end{table}
\label{sec:s2orcb}

\subsection{Sample generated summary}
\begin{figure}[h]
\centering
\includegraphics[scale=0.79]{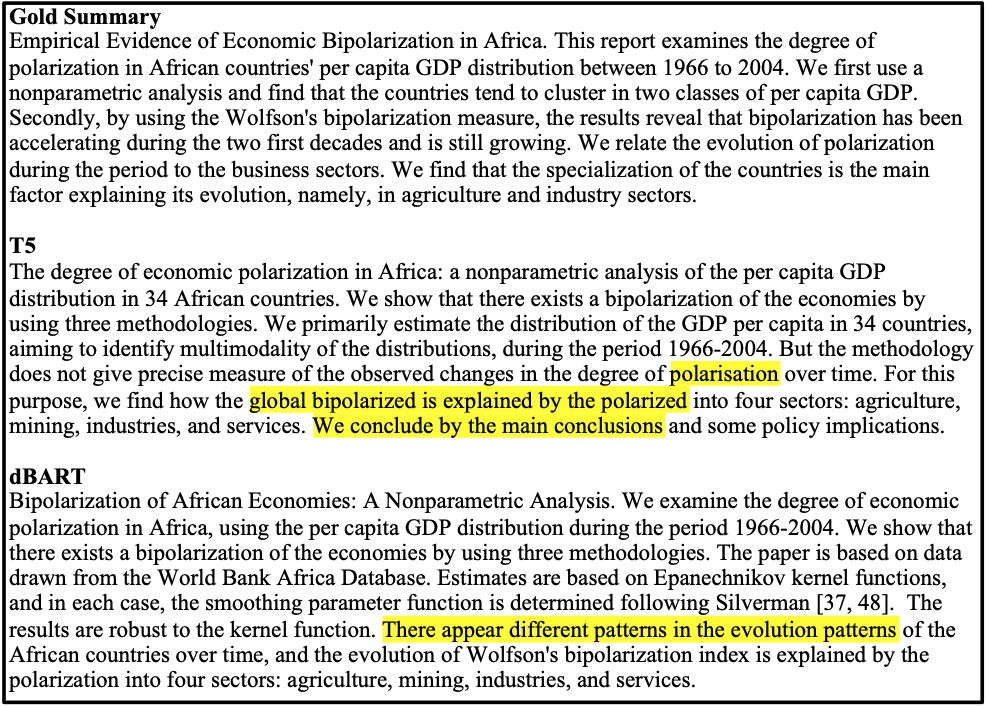}
\caption{Generated summary sample from T5 and dBART with the corresponding Gold (reference) summary. The portions in generated summary indicating grammatical incoherence are highlighted.}
\label{gen_sample}
\end{figure}

We present sample generated summaries from both T5 and dBART along with the gold reference summary in \autoref{gen_sample}. As highlighted in the figure, following instances of grammatical inconsistencies are observed in the generated summaries starting from the top:
\begin{enumerate}[i)]
    \item inconsistent spelling of the term \textit{polarization}
    \item global \textit{bipolarization} is explained by the \textit{polarization}
    \item \textit{We conclude by the main conclusions} seems incoherent
    \item \textit{evolution} is \textbf{not} followed by \textit{patterns} as per the source text
\end{enumerate}

\label{sec:gen_result}
\end{document}